\documentclass[aps,prb,twocolumn,amsmath,amssymb]{revtex4}
\usepackage{graphicx}
\usepackage{dcolumn}
\usepackage{bm}

\usepackage{ulem}
\usepackage{color}
\definecolor{darkred}{RGB}{120,0,0}

\begin{document}\setlength{\unitlength}{1mm}

\title{Exact time-dependent density functional theory for impurity models
}

\author{P. Schmitteckert$^{1,2}$, M. Dzierzawa$^3$ and P. Schwab$^3$}
\affiliation{$^1$ Institute of Nanotechnology, Karlsruhe Institute of Technology, 76021 Eggenstein-Leopoldshafen, Germany}
\affiliation{$^2$ Center of Functional Nanostructures, Karlsruhe Institute of Technology, 76131 Karlsruhe, Germany}
\affiliation{$^3$ Institut f\"ur Physik, Universit\"at Augsburg, 86135 Augsburg, Germany}

\date{\today}

\begin{abstract}
We employ the density matrix renormalization group to construct the exact time-dependent 
exchange correlation potential for an impurity model with an applied transport voltage.
Even for short-ranged interaction we find
an infinitely long-ranged exchange correlation potential which is built up
{instantly} after switching on the voltage.
Our result demonstrates
the fundamental difficulties of transport calculations based on time-dependent density functional theory. 
While formally the approach works, important information can be missing in the ground-state 
functionals and may be hidden in the usually unknown non-equilibrium functionals.
\end{abstract}

\maketitle
\section{Introduction} 
In recent years a combination of Kohn-Sham density functional theory (DFT) and the Landauer approach to transport
has been developed that allows an ab-initio calculation of the current-voltage characteristics of
molecules that are attached to reservoirs.
Early comparisons between calculated and experimental conductances yielded 
discrepancies of several orders of magnitude, while in more recent comparisons a typical discrepancy 
of one order of magnitude was reported.\cite{lindsay2007} 
{Concerning experiments, the reproducibility of $I$-$V$ characteristics in molecular electronics poses a problem,
as the contact configuration may change from sample to sample,
which makes the comparison between theory and experiment difficult.}

First DFT studies of lattice models like the Hubbard model or spinless fermions 
date already back to the late 1980s.\cite{gunnarsson1986,schonhammer1987,schonhammer1995} 
At that time ground state DFT was the main focus of interest. In the 2000s a lattice version of the local density approximation
was constructed using the Bethe ansatz solution of the Hubbard model, 
and the accuracy of the method was tested in some detail.\cite{lima2003,schenk2008,dzierzawa2009}
Due to the mentioned discrepancies between measured and calculated conductances through molecules, 
recently the focus turned to the DFT description of transport.
In the conventional DFT approach to the two-terminal {linear} conductance of a molecular system 
the Kohn-Sham equations are solved in order to obtain the electronic structure, 
and the conductance of the Kohn-Sham system, $G_{\rm KS}$, is {used as a} theoretical estimate 
of the true conductance, $G$. There are thus two possible sources for errors,
(i) the electronic structure calculation and (ii) the identification of $G$ with $G_{\rm KS}$.
While in complex realistic structures it is not possible to quantify (i) and (ii) this is possible
for lattice models, as demonstrated by Schmitteckert and Evers.\cite{schmitteckert2008}
They observed that the Kohn-Sham conductance, $G_{\rm KS}$, based on the 
exact ground state exchange correlation potentials becomes accurate, i.e. $G_{\rm KS}= G$, 
close to (isolated) conductance resonances.
Remarkably this holds even if the spectral properties of the Kohn-Sham system strongly differ from the true ones.\cite{schmitteckert2008,troster2011}
For example, the zero temperature conductance through a Kondo impurity is captured correctly by $G_{\rm KS}$.
The reason for this coincidence is the Friedel 
sum rule which guarantees that $G$ and $G_{\rm KS}$  have identical functional dependencies on 
the charge density.\cite{schmitteckert2008,mera2010,mera2011}
{However, as soon as the conductance is not given by a local Friedel sum rule
there can be orders of magnitude between $G_{\rm KS}$ and the true conductance,\cite{troster2011,bergfield2011,stefanucci2011}
and it can even be parametrically wrong.\cite{PS:DSB2012}}
{It is important to note that this is not a general failure of DFT; rather the assumption that the conductance
of the physical electrons is given by the conductance of the Kohn-Sham particles breaks down.}
It has been stated several times in the literature that this discrepancy can be corrected by including dynamic
contributions to the exchange-correlation potential that are not captured by ground state 
functionals.\cite{stefanucci2004,koentopp2006,schenk2011} 
For the linear current $I$ through an impurity 
this dynamic exchange-correlation potential $V^{\rm xc}$ renormalizes the voltage, such that 
\begin{equation}
\label{eq1}
  I= G V =  G_{\rm KS}(V + V^{ \rm xc} ).
\end{equation}
It is not clear whether a similar voltage renormalization is also present in the non-linear response.

{The goal of our study is to investigate the nature of the dynamical response within a DFT framework beyond the linear transport regime.}
To this end we construct the exact {time-dependent} exchange-correlation potential 
for an impurity model with an applied transport voltage.
{Specifically we construct the time-dependent Kohn-Sham potentials for a two-terminal transport setup, which by
construction yields the correct physical current within the time-dependent DFT description.}
In this paper we will concentrate on a one-dimensional lattice model,
where we can directly compare {time-dependent} DFT with accurate numerical 
and analytical results obtained by many-body techniques. {\cite{boulat2008}}
In order to obtain meaningful results we have to solve much larger systems (here: $240$ lattice sites) 
than for instance in Ref. \onlinecite{verdozzi2008} ($6$ to $12$ lattice sites). In addition the densities have to be calculated 
with an accuracy
better than $ \sim 10^{-6}$; otherwise the reverse engineering of the time-dependent exchange-correlation potentials fails.

We consider the interacting resonant level model, i.e. a one-dimensional model of spinless fermions, where a single interacting
level is coupled to a left and a right lead,
\begin{eqnarray}
H &=& H_\mathrm{L} + H_\mathrm{LR}+ H_\mathrm{R}  \label{eq2} \\ 
H_\mathrm{L}&=& - t \sum_{i=1}^{N_\mathrm{L}-1} c_{\mathrm{L} i}^+ c^{}_{\mathrm{L} i+1} + {\text{h.c}.} \\
H_\mathrm{R}&=& - t \sum_{i=1}^{N_\mathrm{R}-1} c_{\mathrm{R} i}^+ c^{}_{\mathrm{R} i+1} + {\text{h.c}.} \\
H_\mathrm{LR}&=&  -t'(c^+ c_{\mathrm{ L},1} +  c^+ c^{}_{\mathrm{R},1} + {\text{h.c.}} ) \nonumber \\
      && + U\left(n-1/2\right)\left( n_{\mathrm{ L},1} + n_{\mathrm{R},1} - 1 \right) 
.\end{eqnarray}
Here $t=1$ is the hopping amplitude, $N_\mathrm{L}$ and $N_\mathrm{R}$ are the numbers of sites in the left and right lead, and $U$ is the
interaction on the contact link.
All the data presented in this article have been obtained for the half-filled lattice model.
We assume that at time $T<0$ the system is in the ground state. At $T=0$ we include a voltage drop
by applying a potential $eV/2$ ($-eV/2$) on the left (right) lead with a linear
crossover on a scale of several sites left and right of the impurity. Then we follow
the time evolution of the system. On time scales that are shorter than the transit time $T_t = L_{\rm lead}/v_F $  
($v_F = 2t$ is the Fermi velocity)
the finite leads act as reservoirs, such that a time-dependent simulation allows to extract the 
current-voltage relation corresponding to infinite leads, see Refs.~\onlinecite{boulat2008} and \onlinecite{PS:Ann2010} 
for details. 
{Note that in our model the  linear conductance is given by a Friedel sum rule.\cite{Boulat_Saleur:PRB2008,schmitteckert2008}
Therefore, $V^{ \rm xc}$ vanishes in the linear regime.}

We consider a Hamiltonian $H$ with parameters $U$ and $t'$ and a Kohn-Sham Hamiltonian $H_{\rm KS}$  
with the same structure as Eq.~(\ref{eq2})  using different parameters $U_{\rm KS}=0$ and
$t'_{\rm KS}$.
The essence of DFT is a one-to-one correspondence between local densities and potentials:  
Starting with an uniquely defined initial state there exists an
-- up to a gauge transformation -- unique set of potentials  $ \{ v_{\rm KS}(T) \}$
such that the
time-dependent densities $\{ n(T) \}$ are identical for $H$ and $H_{\rm KS}$.
The numerical task is to calculate first the time-dependent local density for $H$ and in the second step the set of potentials 
$\{ v_{\rm KS} \}$ for $H_{\rm KS}$. 
We fix the gauge by imposing that the sum of the potentials is zero.
To calculate the particle density we follow two strategies: 
(a) The particle density in an interacting system is obtained through time-dependent density matrix renormalization group (td-DMRG)
as described in Refs. \onlinecite{PS:Ann2010} and \onlinecite{PS:PRB2004}.
Since these calculations are extremely time-consuming we also 
(b) study a toy model of non-interacting fermions,
where the time-dependent density can be obtained straightforwardly by exact diagonalization.
As mentioned before, when calculating $\{ v_{\rm KS} \}$ from the densities a high accuracy is necessary. We use an iterative procedure that
stops when the densities in $H$ and in $H_{\rm KS}$ agree within an error of $10^{-10}$.

\section{Noninteracting toy model}
\begin{figure}
\includegraphics[width=0.40\textwidth]{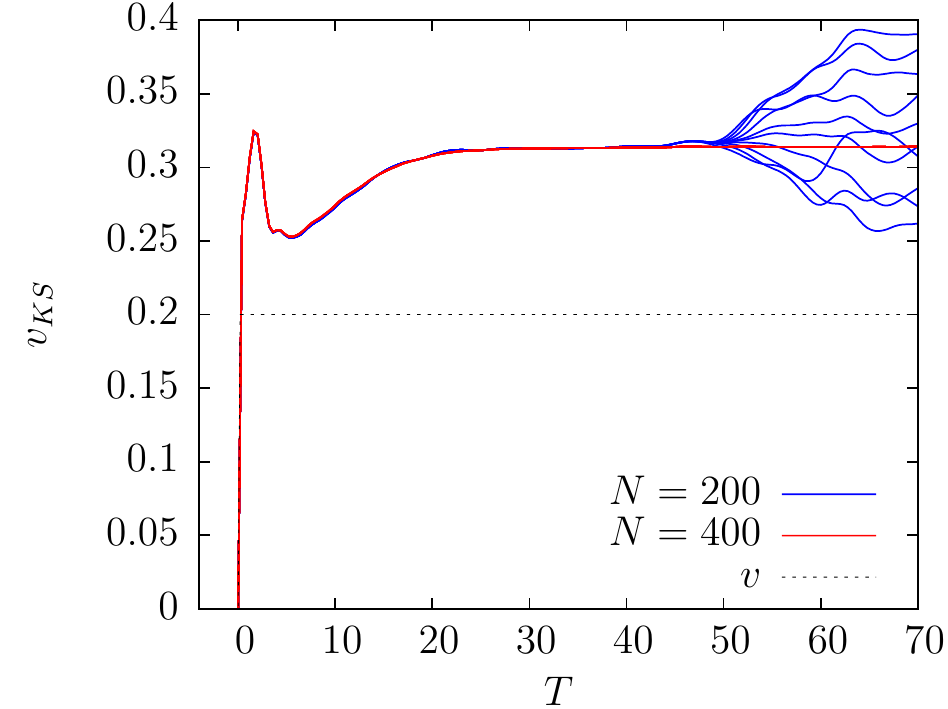}
\caption{ 
\label{fig1}
The Kohn-Sham potential $v_{\rm KS}$ as a function of time for the first ten lattice sites from the left edge of the chain.
The two data sets are for a total chain length 
$N=200$ and $N=400$. 
For comparison the potential $v=0.2t$ in the toy model (corresponding to a voltage $eV=0.4t$) is also shown as dotted line.
}
\end{figure}

We start by considering the non-interacting version of the Hamiltonian (\ref{eq2}), i.e. we set the interaction $U$ as well as 
$U_{\rm KS}$ to zero. 
There remains only one free parameter, the hopping amplitude $t'$ between the impurity site and the leads. 
In our toy model we choose $t' = 0.5t$ and $t'_{\rm KS}=0.3t$.
Typical results for the time-dependent Kohn-Sham potentials are shown in Fig.~\ref{fig1}.
In the toy model a potential $eV=0.4t$ corresponding to a potential $v=\pm 0.2t$ in the left (right) lead is switched on at $T=0$.
As a response 
a current flows through the resonant level that becomes stationary after $T \approx 30$ (the time is in units $\hbar/t$). 
By definition, the same current has to flow also in the Kohn-Sham system; here we find initially a time-dependent voltage 
which becomes stationary {after} $T \approx 30$.
Remarkably the exchange-correlation potential (the difference between $v_{\rm KS}$ and $v$) 
is nearly position independent in the leads, and appears immediately after switching on the voltage.
In the figure, data for a chain of 200 lattice sites and a chain of 400 lattice sites are presented.
In both cases we plot the potentials for the first ten sites at the left edge of the chain.
For the longer chain $v_{\rm KS}$ remains stationary until the end of the simulation, 
and only a single line is seen meaning that the potential is homogeneous in space.
In the shorter chain the potential becomes position and time-dependent after
$T \approx 50$.  
This happens since the simulation time is longer than the transit time $T_t = 100/2 =50$. 

\begin{figure}
\includegraphics[width=0.40\textwidth]{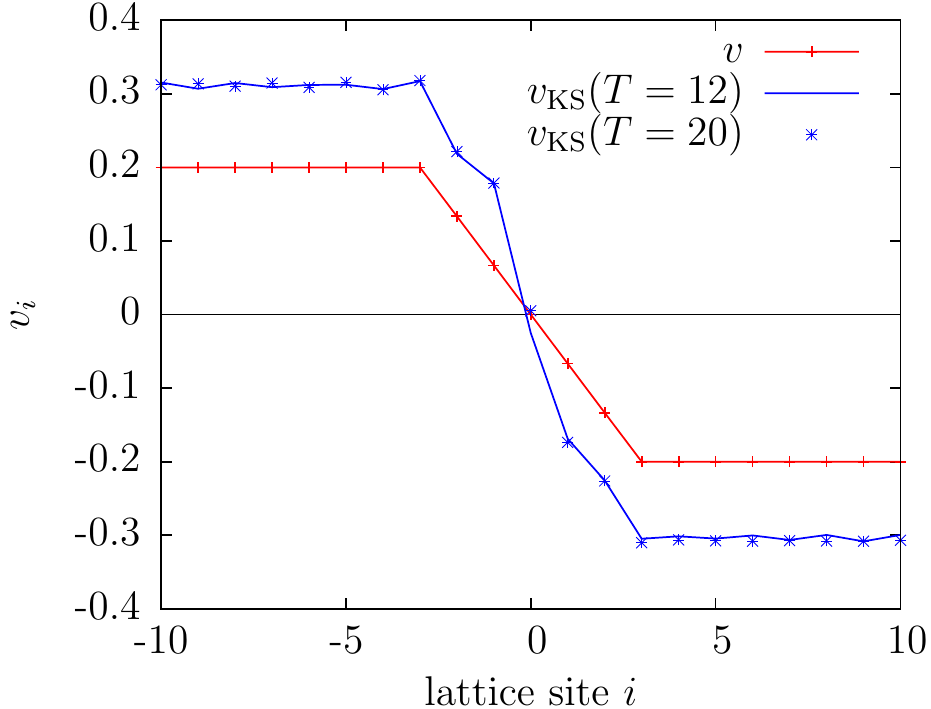}
\caption{\label{fig2}
Local potential $v_i$ in the toy model and in the Kohn-Sham system
close to the impurity ($i = 0$) at time $T = 12$ (solid curve) and $T=20$ (symbols).
}
\end{figure}

\begin{figure}
\includegraphics[width=0.40\textwidth]{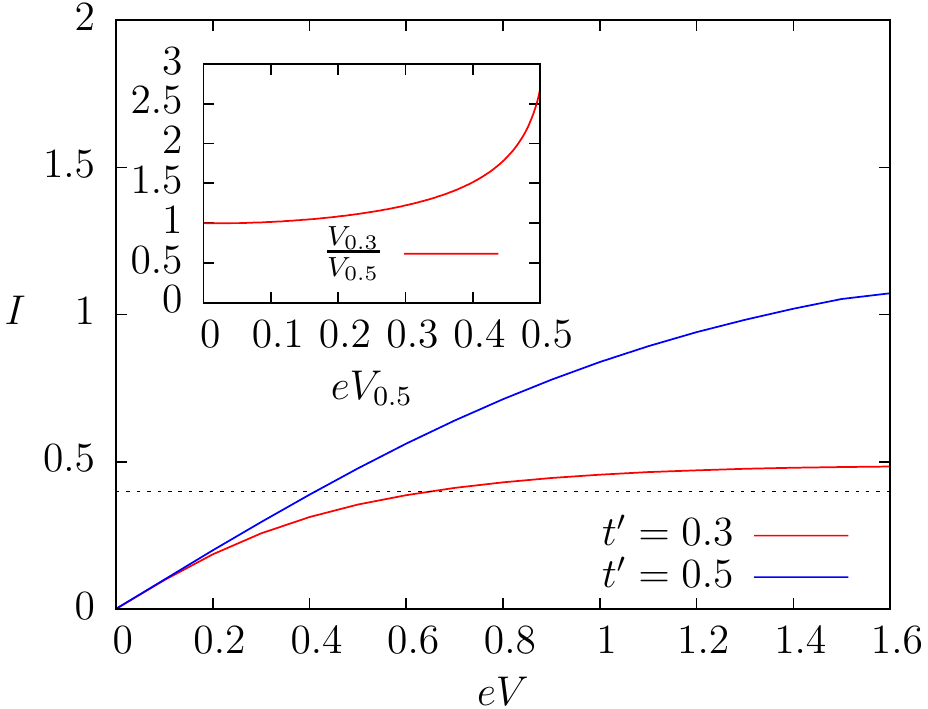}
\caption{\label{fig3} Current (in units of $e t/h$)
as a function of voltage for two non-interacting systems with $t'= 0.3t$ and $t'=0.5t$, respectively. 
In order to generate the same current in both systems (e.g. $I = 0.4$ as indicated by the dotted line) different potentials, $eV_{0.3}$ and $eV_{0.5}$, have to be applied.
The voltage renormalization $V_{0.3}/V_{0.5}$ displayed in the inset gives a reasonable estimate 
of the voltage renormalization observed in the Kohn-Sham Hamiltonian for the toy model.
}
\end{figure}

The Kohn-Sham potential as a function of position is illustrated in Fig.~\ref{fig2}, demonstrating that a major effect of the
exchange-correlation potential is a time-independent renormalization of the {local potentials, which implies an additional
voltage as anticipated in Eq.~(\ref{eq1}).}
A reasonable estimate for the voltage renormalization is obtained
by comparing the $I$-$V$ characteristics for two noninteracting models with $t'=0.3t$ and $t'=0.5t$, see Fig.~\ref{fig3}. 
The current $I=0.3883$ (in units of $et/h$) corresponds to $eV=0.4t$ for $t'=0.5t$ and $eV=0.6061t$ for $t'=0.3t$
which is close {to} the voltage we found in the Kohn-Sham system,
see Fig.~\ref{fig1}.
Notice that the maximum current that can be achieved is larger in the case $t'=0.5t$ than for $t'=0.3t$.
For DFT this implies that no stationary Kohn-Sham potential can generate such large currents. Theoretically, time-dependent
Kohn-Sham potentials could generate a stationary current, however, the proofs of existence of time-dependent Kohn-Sham potentials 
\cite{runge1984,leeuwen1999,ruggenthaler2011} 
do not apply to the present situation, so that possibly the high voltages are not $v$-representable, compare 
Refs. \onlinecite{baer2008} and \onlinecite{li2008}.
A similar situation will also arise in the next section where we investigate the model with interaction.

\section{DMRG results for the interacting model}
We now turn our attention to the interacting case and choose $H$ with
$U= 2 t $ and $t' = 0.3t$, whereas for the Kohn-Sham Hamiltonian we set $U_{\rm KS}=0$ and $t'_{\rm KS}=0.3t$.
The $I$-$V$ characteristics
of this model is known analytically\cite{boulat2008} from field theoretical methods.
For example, the current is given in closed form by\cite{PS:PRL2011}
\begin{equation} \label{eq6}
    I(V) = \frac{e^2 V }{2\pi \hbar } 
    \;{}_3F_2 \left[ \left\{ \frac{1}{4},\frac{3}{4}, 1 \right\}, 
                     \left\{ \frac{5}{6},\frac{7}{6}     \right\}, - \left( \frac{V}{V_{\mathrm c}}\right)^6   \right],
\end{equation}
where $\;{}_3F_2$ is a hypergeometric function and $e V_{\mathrm c} = r {t'}^{4/3} $ 
sets the scale separating a charge $2e$ dominated low voltage
regime, and a charge $e/2$ dominated high voltage regime.\cite{PS:PRL2010}
The regularization $r\approx 3.2$ was determined from 
numerical simulation.\cite{boulat2008,PS:PRL2010,PS:PRL2011}
The study of the toy model in the previous section shows that in order to obtain meaningful results
the simulation time should be of the order $T \approx 30$ or longer, 
so that the length of each lead should be at least $ L_{\rm lead} \approx 60$ sites.
{This means that the excitations originating form the charge quench at the impurity 
should not reach the boundary of the leads within the simulation time.}
{In order to obtain  the time-dependent densities we apply the full td-DMRG \cite{PS:PRB2004} 
as it has already proven to provide accurate results that agree perfectly with the analytic solution.}

{The quench with a symmetric charge imbalance leads to oscillations with frequency $eV/2$ during the transient time and 
finite size induced Josephson like oscillations with frequency $eV$ in the 
steady state regime.\cite{PS:Ann2010} Therefore, when the effective voltage of the Kohn-Sham system is
different from the voltage of the physical one, the time-dependent potentials have to compensate this mismatch. In order to reduce
these finite time and finite size effects we ramp up the voltage linearly from $T=0\ldots5$, leading to reduced
amplitudes of the transient oscillations.}
\begin{figure}
\includegraphics[width=0.45\textwidth]{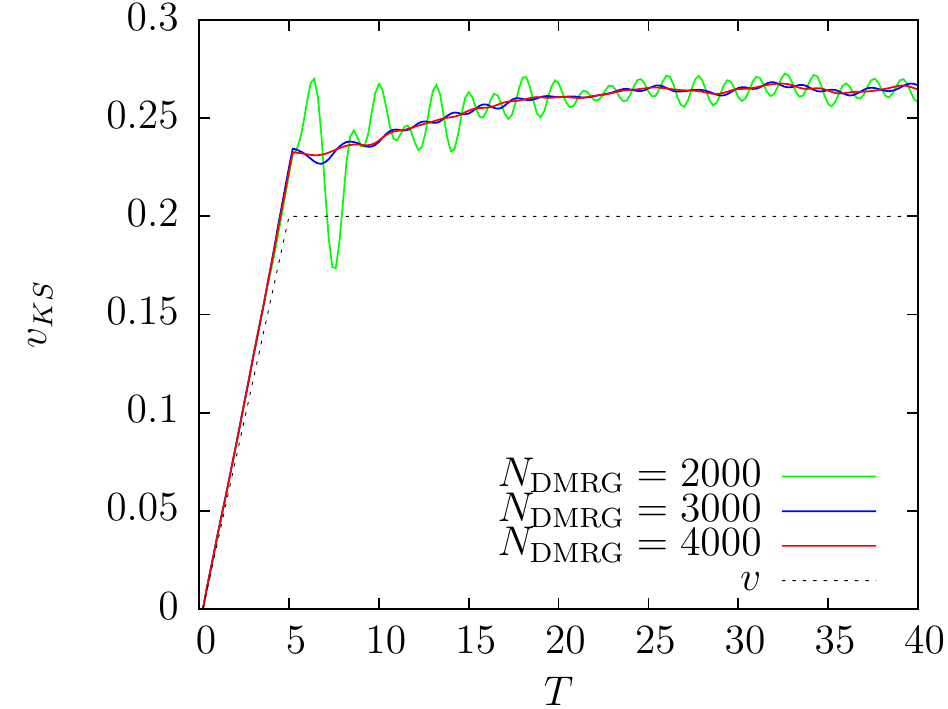}
\caption{\label{fig4} Local potentials on the first ten sites of the left lead as a function of time for an interacting system
($U=2t$, 240 lattice sites). Here we smoothly switched on the external voltage $V = 0.4t$ 
between $T=0$ and $T=5$.
The number of states per block kept in the DMRG calculations varied between 2000 and 4000.
}
\end{figure}
Fig.~\ref{fig4} shows the Kohn-Sham potential in the left lead. 
As in the non-interacting toy model
the main effect is a voltage renormalization.

To obtain numerically converged data is a hard task.
The figure shows results from three different DMRG runs varying the numerical accuracy. In all three runs the time-dependent
particle density is almost identical, variations occur on the scale $10^{-6}$. The Kohn-Sham potentials 
{turn out to be}
very sensitive functions of the particle density and show pronounced differences in the three cases. Only for a very large number
of states per block in the DMRG calculations (highest accuracy) the final result becomes smooth.
\begin{figure}
\includegraphics[width=0.40\textwidth]{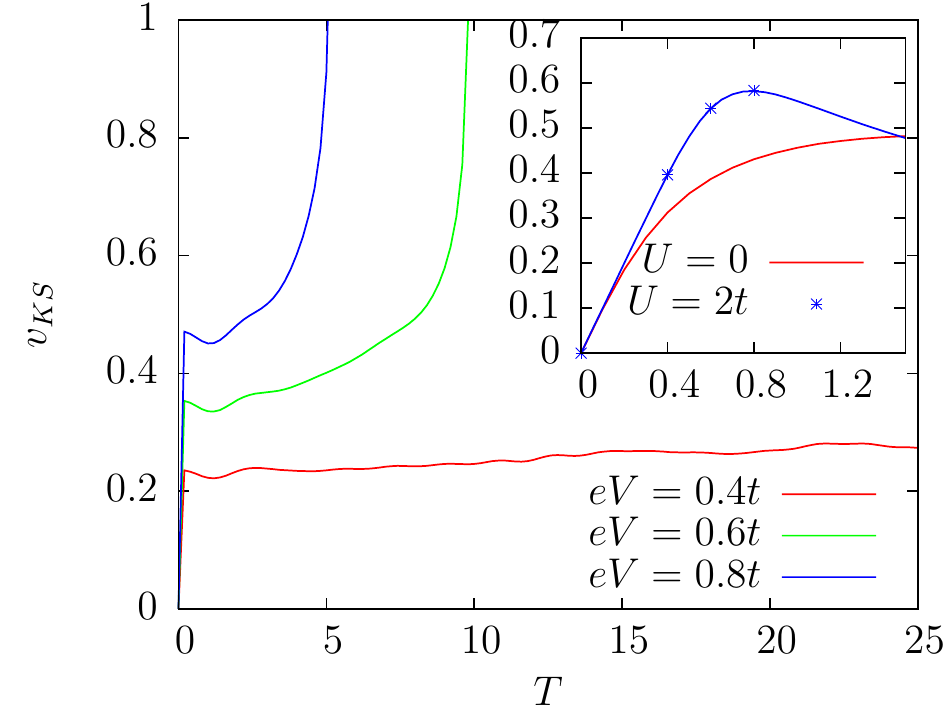}
\caption{\label{fig5}Time-dependent Kohn-Sham potentials on the first ten sites of the left lead
varying the voltage applied to the system {($t'=0.3t$ and $U=2t$)}.
While for a small voltage ($eV =0.4t$) we are able to find a set of Kohn-Sham potentials
for all times, for larger voltages ($eV = 0.6t$, $eV = 0.8t$)
this is only possible for short times.  
The inset shows the $I$-$V$ characteristics for $U=0$ and $U=2t$. 
The curve for $U=2t$ is given by Eq. (\ref{eq6}) and the symbols are the DMRG results.
}
\end{figure}
In Fig.~\ref{fig5} time-dependent Kohn-Sham potentials for different voltages are depicted. 
Whereas for a small voltage we are able to find Kohn-Sham
potentials for all times, this is not the case for large voltage where the Kohn-Sham potential diverges at finite $T$.
It will be interesting to see in future research, whether the time scale of the singularity has a deeper meaning.
Since the $v$-representability is not guaranteed in the lattice model, we believe that Kohn-Sham potentials do not exist for all times
in these cases, compare also the discussion relating to Fig.~\ref{fig3}. 

{In the inset of Fig.~\ref{fig5} we show the $I$-$V$ characteristics of the noninteracting ($U=0$) 
and interacting ($U=2t$) resonant level model.
Again, there is a regime, where the current in the interacting case is higher than any current achievable in the noninteracting case.
Since the noninteracting case is similar to the Kohn-Sham system, this is a hint that 
no stationary Kohn-Sham potential might exist in this regime.
The cases where we find diverging Kohn-Sham potentials are indeed in the regime where the current for $U=2t$ exceeds the one for $U=0$.}
Of course, the Kohn-Sham systems at finite voltage have additional potentials with a spatial structure which is not captured by
a single number, $V^{\mathrm{xc}}$.
It is also important to note, that the Kohn-Sham potentials
should not create a current outside the ``light cone'' $v_F T$. 
The continuity equation then implies that any oscillations of the potentials
outside this regime have to be instantaneous and constant in space including the reservoirs.
{While one can gauge the resulting $V^{\mathrm{xc}}$ into a time-dependent phase 
$\exp( \mathrm{i}V^{\mathrm{xc}}T)$ of a hopping element
    at the impurity, one loses the property of an at least locally stationary system found in this work.}

\section{Summary}
We calculated the exchange-correlation potential for a one-dimensional model system with an applied transport voltage.
Specifically we considered an impurity model with short-ranged interaction and non-interacting leads.
Immediately after switching on the voltage the exchange-correlation potential appears deep inside the leads.
This long-ranged potential is purely dynamic, and therefore approximations based 
on equilibrium functionals and short-ranged approximations have to fail.

This work has been supported by the Deutsche Forschungsgemeinschaft through TRR80.
We would like to thank Ferdinand Evers, Claudio Verdozzi, and Peter W{\"o}lfle for insightful discussions.

\end{document}